# Coronary artery calcification assessment in National Lung Screening Trial CT images (DeepCAC2)

Leonard Nürnberg[1,2,3,4], Simon Bernatz[1,2,4,5], Borek Foldyna[1,6], Michael T. Lu[1,6], Andrey Fedorov[7], Hugo JWL Aerts[1,2,4]

[1] Artificial Intelligence in Medicine (AIM) Program, Mass General Brigam, Harvard Medical School, Boston, MA, USA.
[2] Radiology and Nuclear Medicine, CARIM & GROW, Maastricht University, Maastricht, the Netherlands.
[3] Department of Radiation Oncology (Maastro Clinic), Maastricht, the Netherlands
[4] Department of Radiation Oncology, Dana-Farber Cancer Institute, Brigham and Women's Hospital, Harvard Medical School, Boston, MA, USA.
[5] Department of Radiology and Nuclear Medicine, Goethe University, Frankfurt am Main, Germany.
[6] Cardiovascular Imaging Research Center, Massachusetts General Hospital and Harvard Medical School, Boston, MA, USA.
[7] Department of Radiology, Brigham and Women's Hospital, Harvard Medical School, Boston, MA, USA.

**ABSTRACT**

Coronary artery calcification (CAC) is a strong predictor of cardiovascular risk but remains underutilized in clinical routine thoracic imaging due to the need for dedicated imaging protocols and manual annotation. We present DeepCAC2, a publicly available dataset containing automated CAC segmentations, coronary artery calcium scores, and derived risk categories generated from low-dose chest CT scans of the National Lung Screening Trial (NLST). Using a fully automated deep learning pipeline trained on expert-annotated cardiac CT data, we processed 127,776 CT scans from 26,228 individuals and generated standardized CAC segmentations and risk estimates for each acquisition. We already provide a public dashboard as a simple tool to visually inspect a random subset of 200 NLST patients of the dataset. The dataset will be released with DICOM-compatible segmentation objects and structured metadata to support reproducible downstream analysis. The deep learning pipeline will be made publicly available as a DICOM-compatible MHub.ai container. DeepCAC2 provides a transparent, large-scale, public, fully reproducible resource for research in cardiovascular risk assessment, opportunistic screening, and imaging biomarker development.

# BACKGROUND & SUMMARY

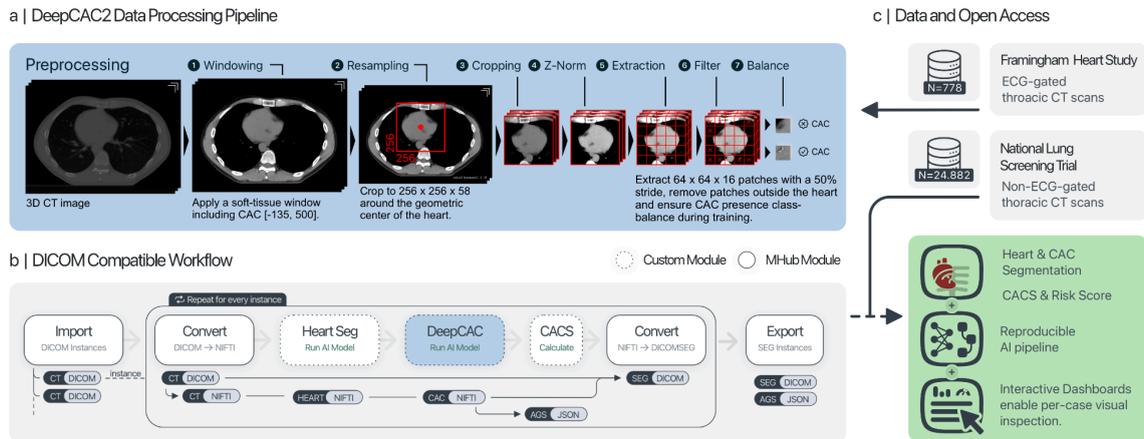

**Figure 1** Public Access Deep Learning Pipeline
a) Preprocessing includes windowing, resampling, cropping around the heart center, z-normalization, patch extraction, patch filtering, and balancing to equal amounts of CAC and non-CAC patches. b) The trained model is embedded in an end-to-end open-source DICOM-compatible pipeline. c) The model was trained using 778 ECG-gated chest CT scans from the Framingham Heart Study. CAC segmentations and AGS scores were generated for 24882 images from the National Lung Screening Trial and are publicly available.

Coronary artery calcification (CAC) is a well-established and robust predictor of cardiovascular disease (Agatston et al. 1990). Although CAC quantification is rapid and largely semi-automatic on dedicated ECG-gated cardiac computed tomography (CT) scans, its routine clinical use remains limited by the need for dedicated acquisition protocols, which are not consistently performed in patients undergoing chest imaging for other indications. This limits the widespread use of CAC scoring, despite the potential for opportunistic assessment from existing chest CT data.

The National Lung Screening Trial (NLST) dataset is a large, publicly available collection of non ECG-gated low-dose chest CT scans acquired as part of a randomized lung cancer screening study (National Lung Screening Trial Research Team, 2011). It contains 203,087 CT scans from 26,254 individuals aged 55 to 74 years with a history of heavy smoking. The dataset includes a baseline scan and up to two annual follow-up scans per participant, enabling longitudinal analyses. Imaging was performed at multiple sites participating in either the Lung Screening Study (LSS) or the American College of Radiology Imaging Network (ACRIN). The dataset shows substantial heterogeneity in scanner vendors, reconstruction settings, and image characteristics. However, imaging was not performed using a dedicated CAC acquisition protocol preventing established CAC quantification.

Importantly, availability of the NLST collection in National Cancer Institute Imaging Data Commons (IDC) (Fedorov et al., 2023) enables continuous enrichment of the images with accompanying annotations (as demonstrated by the recent additions of the Sybil (Krishnaswamy et al., 2025b; Mikhael et al., 2023) and NLSTSeg (Chen et al., 2025; Krishnaswamy et al., 2025a) tumor annotations), segmentations and analysis results, lowering barriers for reuse of this data. NLST CAC annotations can further enrich this dataset and enable new secondary studies.

We have previously demonstrated the clinical value of a deep learning system (DeepCAC) for the automatic quantification of CAC and prediction of cardiovascular events in lung cancer screening CT (Zeleznik et al., 2021). However, recent developments in AI motivated us to revisit the earlier developed approach, while at the same time improving analysis transparency and reproducibility. The publicly available dataset presented here enables new directions for secondary use of the NLST images, while empowering the users of the DeepCAC2 model to scrutinize our results and develop confidence in the model prior to using it.

Therefore, we developed a fully automated and reproducible deep learning pipeline for CAC segmentation and quantification and applied it to the entire NLST dataset. The segmentation model was trained and evaluated on 778 high-quality chest CT scans from the Framingham Heart Study (FHS) with expert CAC annotations (D'Agostino et al., 2008). The model uses a 3D U-Net (Azad et al., 2024; Ronneberger et al., 2015) architecture trained on overlapping 3D patches, with preprocessing that restricts training to the cardiac region and balances sampling by equally including CAC-positive and CAC-negative patches. Predictions are aggregated at the voxel level from the overlapping predictions to generate volumetric CAC segmentations.

From these segmentations, coronary artery calcium scores (CACS) and categorical CAC risk groups are derived using established clinical thresholds. Using this pipeline, we processed 127,776 CT scans from 26,228 NLST participants, covering all available time points and reconstruction kernels. For each scan, we provide DeepCAC2-generated CAC segmentations, CACS derived from those segmentations, and the corresponding risk classifications. These derived annotations enable systematic incorporation of CAC as a variable in NLST-based analyses, supporting stratification by CAC presence and burden, longitudinal and survival analyses, and targeted subcohort selection. In particular, they enable direct filtering of patients with and without CAC, facilitating studies focused on specific cardiovascular risk profiles. By providing standardized CAC measurements across the full cohort, this work extends the utility of NLST for reproducible research in cardiovascular risk assessment and opportunistic screening using large-scale chest CT data.

**METHODS**

*Deep Learning Pipeline*

For training the AI pipeline, we used n = 778 chest CT scans from the FHS dataset. CAC was manually annotated in all CT images by expert CT readers. Cardiac segmentations were created for all scans using the *platipy* nn-UNet based cardiac segmentation model (Finnegan et al., 2023; Isensee et al., 2021). We identified the presence of CAC in the coronary arteries for each sample. The dataset was then split into training and validation in an 80/20 ratio, maintaining an even distribution of patients with CAC presence. This resulted in n = 622 CT images for model training and n = 156 for model validation.

We first resampled all scans to a uniform spacing of 0.7 x 0.7 mm and a uniform slice thickness of 2.5 mm to ensure spatial consistency. We cropped the volume to a bounding box of 256 x 256 x 58 voxels, centered on the heart, with dimensions selected to accommodate the largest heart in the training dataset. We applied a soft-tissue window of [-135, 500] hounsfield units (HU) with an increased upper bound to capture calcifications. Each scan was subjected to z-normalization to standardize the intensity distributions, followed by linear rescaling to the range [0, 1] to optimize the stability of the network. To efficiently train the model while preserving spatial context, we extracted overlapping 3D patches of size 64 x 64 x 16 voxels with a step size of 32 x 32 x 8 voxels, achieving 50% overlap in all three dimensions. The class imbalance between patches containing CAC (n = 120,077) and patches absent of CAC (n = 29,713) was addressed by randomly sampling CAC-free patches from across all pre-computed patches. This process resulted in an even 50/50 split of empty and CAC containing patches and generated 37,534 patches for training and 9,034 for validation.

We then trained a 3D U-Net to predict the presence of CAC per voxel using backpropagation. The loss between the predictions and the ground truth masks was calculated using BinaryDiceLoss (Milletari et al., 2016). The learning rate was set to 0.0001. Adam optimizer and gradient scaling were used. The model was trained with a batch size of n=32 concurrent patches. Training was performed on a single Quadro RTX 8000 GPU for 50 epochs and completed after 8 hours and 54 min.

The model predictions were aggregated patchwise by averaging results across overlapping patches, followed by applying a threshold of 0.5 to classify voxels as CAC or non-CAC. CACS were calculated from the predicted CAC segmentation masks and converted into categorical CAC risk groups using established clinical thresholds. Risk group 0 (very low) corresponds to a CACS of 0, risk group 1 (low) to scores between 1 and 100, risk group 2 (moderate) to scores between 101 and 300, and risk group 3 (high) to scores greater than 300.

*Model Evaluation*

To assess the *technical validity* of the CAC segmentations and derived CACS, we compared automatically computed scores with expert annotations available for a subset of NLST scans (Zeleznik et al., 2021). This independent test set included 390 CT scans with expert CAC segmentations generated using the open-source software 3D Slicer V4 (Pieper et al., 2005). Agreement between automated and expert-derived CACS was evaluated using Spearman correlation, and agreement in categorical CAC risk groups was assessed using a weighted Cohen's kappa statistic.

To assess the *clinical validity* of the derived CACS and CAC risk stratification, survival analysis was performed on the NLST cohort (n = 23011). Because the released dataset contains multiple scans per participant across timepoints, a single baseline CT scan per individual was selected to enable patient-level survival analysis. Only baseline (T0) scans were included, and images were restricted to AXIAL/ORIGINAL/PRIMARY acquisitions with slice thickness ≤3.0 mm and reconstruction kernels (STANDARD, B50f, FC51, or C) to ensure consistent image quality. If, after these filters, multiple scans were available for a participant, we selected the scan with the largest number of slices to maintain a single observation per individual. Overall survival was defined using all-cause mortality as the event. Time-to-event was calculated from baseline screening to death or last follow-up

(censoring). Kaplan-Meier survival analysis was used to estimate overall survival across CAC risk groups derived from the automatically computed CACS. Multivariable Cox proportional hazards regression was performed to evaluate the association between CAC risk group and all-cause mortality, adjusting for age, sex, smoking status, and history of heart disease. Model discrimination was quantified using the concordance index.

*Model Distribution*

We adapted the MHub.ai (Nürnberg et al., 2026) framework to package our AI pipeline into a self-contained, customizable, reproducible and end-to-end DICOM compatible processing pipeline. MHub.ai is a platform that aims to standardize and simplify AI models for medical imaging. The [MHub.ai](#) framework provides a set of utility modules for reading and generating DICOM files, file format conversion, numeric output reporting, file organization and more. The framework can also be extended with custom modules that integrate seamlessly with existing modules. For our pipeline, we used the following configuration of modules, which are executed sequentially according to the standard workflow: The *DicomImporter* module reads DICOM files and automatically organizes them into series for sequential processing. The *NiftiConverter* module is then used to convert the images into the format required by our pipeline. To create the heart segmentation, we use the NNUnerRunner module and add pre-trained weights for heart segmentation (Finnegan et al., 2023). We developed two custom MHub modules, the *DeepCACRunner* to run the *DeepCAC2* AI model pipeline and the *AGSCalculator* module that calculates the CACS from the CT image and the CAC segmentation. We then use the *DSegConverter* module to generate compliant DICOMSEG files for the CAC segmentations and a *ReportExporter* module to export the CACS and risk assessment to a customizable JSON format. The container image includes the inference pipeline and all model weights in compliance with the [MHub.ai](#) requirements.

*Data Generation*

We obtained a local copy of 128,037 CT scans from 26,229 NLST participants from the Imaging Data Commons (IDC) repository (Fedorov et al., 2023) by querying the publicly available NLST collection and applying predefined selection criteria that included slice thickness between 1 and 5 mm, more than 50 slices per series and consistent slice geometry (e.g., single slice per spatial position, no missing slices, consistent resolution and slice dimensions, see selection query provided in Appendix). The selected images were processed sequentially using our deep learning pipeline on four Quadro RTX 8000 GPUs in parallel. For each CT scan, we generated volumetric segmentations of the heart and auto-detected CAC, saved as DICOM Segmentation (SEG) objects (Fedorov et al., 2016; Herz et al., 2017), and a JSON file referencing the CT and SEG series by the corresponding SeriesInstanceUIDs for provenance, along with the calculated CACS and risk classification. For 261 images (0.2%), processing errors occurred and no segmentations were generated, affecting a single patient. Processing was completed in about 6 days.

Model inference was performed using the MHub.ai Docker image, which provides a standardized execution environment. The same container can be used to reproduce the segmentations reported here or to generate CAC annotations for additional CT data using the following command:

```
export input = /path/to/input/ct/scan/dicom/files/
export output = /path/to/output/directory/
docker run –rm -it -v $input:/app/data/input_data:ro -v
$output:/app/data/output_data mhubai/deepcac:latest
```

**DATA RECORD**

The dataset includes the segmentation of the heart and CAC, the computed CACS, and the derived risk for 26,228 subjects across a total of 127,776 CT scans. The segmentations are provided in DICOM SEG format, which contains both the heart and CAC segmentations predicted by our deep learning algorithm. The CACS and risk score are provided as key-value pairs in a complementary JSON file with three keys: "Predicted Agatston Score", which stores the CACS value; "Predicted Risk Group", which contains the classified risk; and "SeriesInstanceUID", which links to the original CT scan series. The data is organized in a folder structure where the first-level folder is the patient ID, and the second-level folder is the image's series instance UID, which contains the cac.seg.dcm segmentation file and the DeepCAC.report.json file. All files are ZIP compressed.

**TECHNICAL VALIDATION**

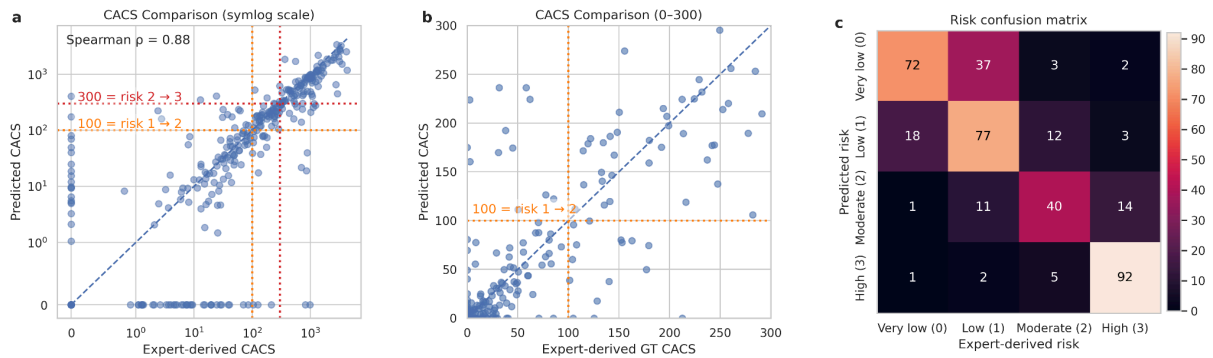

**Figure 2** a) Comparison of predicted and expert coronary artery calcium scores (CACS) of n = 390 CT images from NLST shows high correlation (Spearman p = 0.879). b) Comparison of predicted and expert CACS focusing on the CACS range 0-300 (n = 271). c) Confusion matrix comparing risk categories (very low = 0, low = 1, moderate = 2, high = 3) showing an almost perfect agreement (weighted kappa κ = 0.844).

Expert CAC annotations were previously performed for CT scans (Zeleznik et al., 2021) and used as an independent test set for quality control. We compared the calcium scores derived from AI-generated segmentations with manually measured calcium scores in 390 patients from the NLST. CACS shows a very high (Mukaka, 2012) Spearman correlation (**Figure 2a, b**) and almost perfect agreement (Viera & Garrett, 2005) between automatic and manually calculated risk groups (**Figure 2c**).

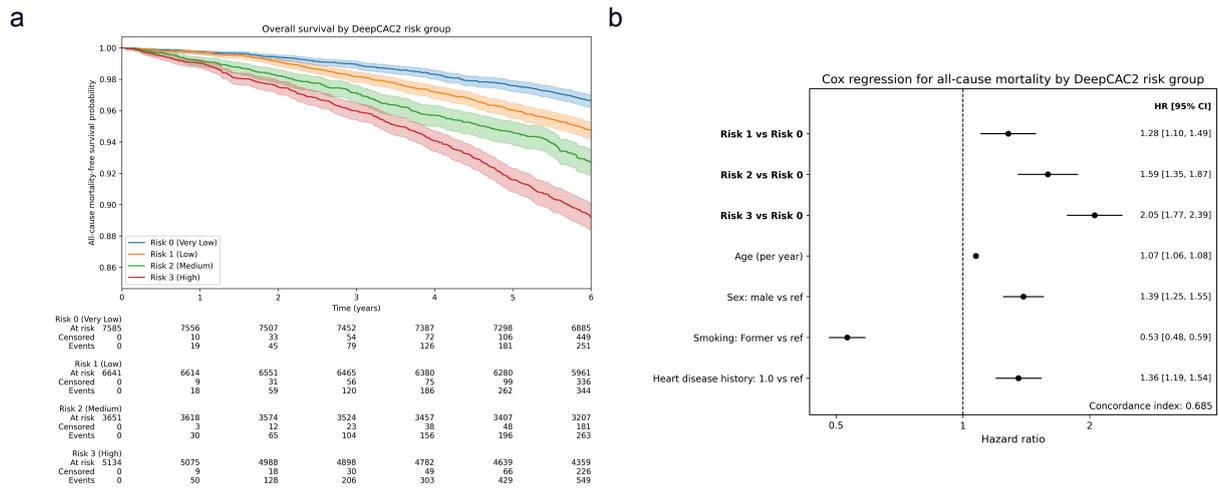

**Figure 3**
a) Kaplan-Meier estimates of overall survival (all-cause mortality-free survival) stratified by DeepCAC2-derived CAC risk group using baseline NLST scans. Risk groups correspond to CACS bins: 0 = 0, 1-100 = 1, 101-300 = 2, >300 = 3. Numbers at risk, events and censored individuals are shown below the plot. b) Adjusted hazard ratios from multivariable Cox proportional hazards regression for all-cause mortality, including DeepCAC2-derived CAC risk group strata and clinical covariates. Hazard ratios are shown with 95% confidence intervals. Risk group 0 is the reference category. Model discrimination is summarized using the concordance index (0.685).

The automatically derived CAC scores showed strong stratification of overall survival in the NLST cohort. Survival analysis included 23,011 participants, with 1,624 deaths and a median follow-up of 6.7 years. Kaplan-Meier analysis demonstrated clear separation of survival curves across CAC risk categories, with progressively lower survival probabilities as CAC burden increased (**Figure 3a**). In multivariable Cox proportional hazards regression adjusted for age, sex, smoking status, and history of heart disease, CAC risk group remained a strong independent predictor of all-cause mortality. Compared with participants with CAC = 0, hazard ratios increased stepwise across risk groups, reaching 2.05 (95% CI 1.77-2.39) for the highest CAC category (**Figure 3b**). Age, male sex, and pre-existing heart disease were also significant predictors of mortality, consistent with established clinical risk factors. The concordance index of the multivariable model was 0.685, indicating moderate to good discriminative performance. These findings confirm the technical validity and clinical relevance of the released CAC scores and risk group assignments.

**USAGE NOTES**

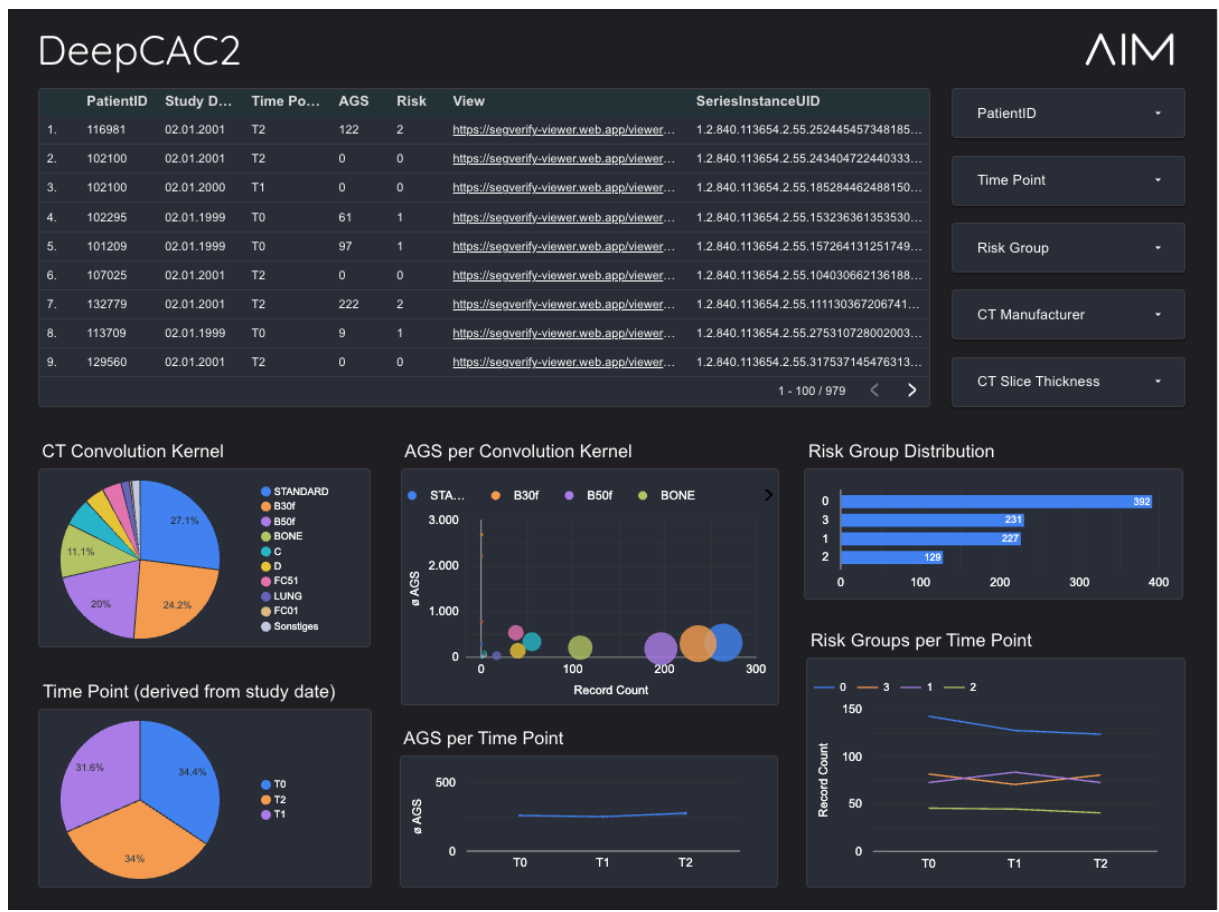

**Figure 4** Interactive dashboard. The top left table provides case-level access to the computed coronary artery calcium scores (CACS), derived risk group, and the CT image with CAC segmentation overlay, which opens in an online OHIF viewer. On the right side, the dashboard can be filtered by patient, timepoint, risk group, manufacturer, or slice thickness. The charts below show the distribution of convolution kernels and timepoints, the average CAC score per convolution kernel and timepoint, the risk group distribution, and risk group separation across all timepoints for the current selection.

To facilitate transparency, demonstrate the usage of the dataset and the value of the interoperable DICOM representation of the segmentation results, we developed an interactive dashboard (**Figure 4**) that can be used to explore the dataset, subset the cases by various clinical, acquisition and AI-derived attributes and visualize individual CT images along with the DeepCAC2 segmentation results. We used the off-the-shelf components available in Google Cloud, similar to how the data is managed in IDC (Fedorov et al., 2023), for developing this dashboard. The DICOM SEG files included in the dataset were first uploaded to a Google Cloud Storage bucket, followed by ingestion into a Google Healthcare API (GHC) [(Cloud Healthcare API | Google Cloud 2020)](#) DICOM Store. We then used GHC capabilities to export DICOM metadata from the GHC DICOM store into a Google BigQuery [(Google Cloud Platform 2026)](#) table. The latter made it possible to query any of the DICOM attributes via Standard Query Language (SQL) interface. We relied on the public BigQuery tables maintained by IDC to join segmentation metadata with the DICOM metadata of the NLST CT images already available in IDC. To enable lightweight visualization of the data we deployed an instance of OHIF Viewer (Ziegler et al., 2020), which is configured to fetch the

CT series from the IDC-maintained public DICOM store and merge those with the DICOM SEG CAC segmentation series available via a lightweight proxy without authentication. Both of those are retrieved via standard DICOMweb interface. All of those components were then used to enable an interactive dashboard implemented using Google Looker Studio [(Google Cloud Platform 2026)](Google Cloud Platform 2026), as shown in Fig.4 and is publicly available at https://lookerstudio.google.com/reporting/2c656fab-89ce-4ccd-a6a6-6c76ba7ecc51/page/P3qpF.

## DATA AVAILABILITY

Upon publication, the trained model will be released via MHub.ai at https://mhub.ai/models/deepcac2, and the pre-built Docker container will be available on Docker Hub in the MHub.ai repository under mhubai/deepcac2. All DeepCAC2-derived CAC segmentations, the calculated CACS and CAC risk scores will be made available on Zenodo and integrated into IDC.

## CODE AVAILABILITY

The corresponding model implementation and inference pipeline will be available upon publication as part of the MHub.ai models repository under https://github.com/MHubAI/models/tree/main/models/deepcac2.

# APPENDIX

## IDC Query for NLST download

```sql
%%bigquery cohort_df --project=$project_id

WITH
 idc_instances_per_series AS (
 SELECT
   SeriesInstanceUID,
   COUNT(DISTINCT(SOPInstanceUID)) AS num_instances,
   COUNT(DISTINCT(ARRAY_TO_STRING(ImagePositionPatient, "/"))) AS position_count,
   COUNT(DISTINCT(ARRAY_TO_STRING(PixelSpacing, "/"))) AS pixel_spacing_count,
   COUNT(DISTINCT(ARRAY_TO_STRING(ImageOrientationPatient, "/"))) AS orientation_count,
   ANY_VALUE(ARRAY_TO_STRING(ConvolutionKernel, "/")) as convkernel,
   MAX(SAFE_CAST(SliceThickness AS float64)) AS max_SliceThickness,
   MIN(SAFE_CAST(SliceThickness AS float64)) AS min_SliceThickness,
   STRING_AGG(DISTINCT(SAFE_CAST("LOCALIZER" IN UNNEST(ImageType) AS string)), "") AS has_localizer
 FROM
   `bigquery-public-data.idc_current.dicom_all`
 WHERE
   Modality = "CT" and
   access = "Public"
 GROUP BY
   SeriesInstanceUID)
SELECT
 dicom_all.SeriesInstanceUID,
 ANY_VALUE(dicom_all.collection_id) as collection_id,
 ANY_VALUE(dicom_all.PatientID) AS PatientID,
 ANY_VALUE(BodyPartExamined) as BodyPartExamined,
 ANY_VALUE(ContrastBolusAgent) as ContrastBolusAgent,
 ANY_VALUE(ARRAY_TO_STRING(ImageType, "/")) as ImageType,
 ANY_VALUE(idc_instances_per_series.num_instances) AS num_instances,
 ANY_VALUE(StudyInstanceUID) AS StudyInstanceUID,
 ANY_VALUE(dicom_all.ClinicalTrialTimePointID) AS ClinicalTrialTimePointID,
 ANY_VALUE(ARRAY_TO_STRING(dicom_all.PixelSpacing, "/")) as PixelSpacing,
 ANY_VALUE(dicom_all.SliceThickness) as SliceThickness,
 ANY_VALUE(ARRAY_TO_STRING(dicom_all.ConvolutionKernel, "/")) as ConvolutionKernel,
 ANY_VALUE(dicom_all.ReconstructionDiameter) as ReconstructionDiameter,
 ANY_VALUE(dicom_all.Manufacturer) as Manufacturer,
 ANY_VALUE(CONCAT("https://viewer.imaging.datacommons.cancer.gov/viewer/", dicom_all.StudyInstanceUID, "?seriesInstanceUID=", dicom_all.SeriesInstanceUID)) AS idc_url
FROM
 `bigquery-public-data.idc_current.dicom_all` AS dicom_all
JOIN
 idc_instances_per_series
ON
 dicom_all.SeriesInstanceUID = idc_instances_per_series.SeriesInstanceUID
WHERE
 idc_instances_per_series.min_SliceThickness >= 1
 AND idc_instances_per_series.max_SliceThickness <= 5
 AND idc_instances_per_series.num_instances > 50
 AND idc_instances_per_series.num_instances/idc_instances_per_series.position_count = 1
 AND has_localizer = "false"
GROUP BY
 SeriesInstanceUID
ORDER BY
 num_instances DESC
```

**AUTHOR CONTRIBUTIONS**

Conceptualization: L.N., A.F., M.T.L., and H.J.W.L.A. Methodology: L.N., S.B., A.F., B.F., M.T.L., and H.J.W.L.A. Software: L.N. Validation: L.N., S.B., A.F., and H.J.W.L.A. Formal analysis: L.N. and S.B. Resources: A.F. and H.J.W.L.A. Data curation: L.N. and S.B. Writing—original draft: L.N., S.B., A.F., and H.J.W.L.A.. Writing—review and editing: all authors. Visualization: L.N. Supervision: A.F. and H.J.W.L.A. Project administration: A.F. and H.J.W.L.A. Funding acquisition: A.F. and H.J.W.L.A.

**COMPETING INTERESTS**

M.T.L. reports serving on an advisory board for Eli Lilly, outside the submitted work. H.J.W.L.A. reports consulting fees and/or stock from Onc.AI, Love Health, Sphera, Health-AI, Ambient, and AstraZeneca. The other authors declare no competing interests.

**FUNDING**

We thank the National Cancer Institute for collecting and making the data from the NLST accessible, and The Cancer Imaging Archive and the Imaging Data Commons for making this and other imaging collections used for developing our deep learning model available on their platforms. M.T.L. reports research funding to his institution from the American Heart Association, AstraZeneca, Ionis, Johnson & Johnson Innovation, Kowa Pharmaceuticals America, National Academy of Medicine, the National Heart, Lung, and Blood Institute, and the Risk Management Foundation of the Harvard Medical Institutions outside the submitted work. H.J.W.L.A. acknowledges financial support from NIH (HA: NIH-USA U24CA194354, NIH-USA U01CA190234, NIH-USA U01CA209414 and NIH-USA R35CA22052; BHK: NIH-USA K08DE030216-01) and the European Union–European Research Council (HA: 866504). S.B. acknowledges funding from the Deutsche Forschungsgemeinschaft (DFG, German Research Foundation)—502050303. The FHS is funded by a contract from the National, Heart, Lung, and Blood Institute (75N92019D0031, contract number 75N92019D0031). The Imaging Data Commons team has been funded in whole or in part with Federal funds from the NCI, NIH, under task order no. HHSN26110071 under contract no. HHSN261201500003I.